\documentclass[pra,twocolumn,showpacs,floatfix]{revtex4}
\usepackage{graphicx}
\begin{document}

\title{Stability of persistent currents in a Bose-Einstein condensate
confined in a toroidal trap}
\author{M. \"Ogren$^{1,2}$ and G. M. Kavoulakis$^3$}
\affiliation{$^1$ARC Centre of Excellence for Quantum-Atom Optics,
School of Physical Sciences, University of Queensland, Brisbane,
Queensland 4072, Australia \\
$^2$Mathematical Physics, Lund Institute of Technology, P.O.
Box 118, SE-22100 Lund, Sweden\\
$^3$Technological Education Institute of Crete, P.O. Box 1939,
GR-71004, Heraklion, Greece}
\date{\today}

\begin{abstract}

Motivated by recent experiments in Bose-Einstein condensed atoms
that have been confined in toroidal traps, we examine the stability
of persistent currents in such systems. We investigate the extent
that the stability of these currents may be tunable, and the possible
difficulties in their creation and detection.

\end{abstract}
 \maketitle

\section{Introduction}

The recent advances in the physics of cold atoms have allowed
experimentalists to engineer many of the properties of these
systems. Remarkably, the experiments of
Refs.\,\cite{Kurn,Olson} have managed  to trap atoms in
toroidal traps, while in Ref.\,\cite{phil} persistent currents
were created and observed in toroidal traps. Furthermore, in an
older experiment, Ref.\,\cite{critvel} has investigated the
dissipationless flow of an obstacle in an elongated
Bose-Einstein condensate below some critical velocity.

Such simple trapping geometries makes these systems very
appealing mainly for two reasons, one theoretical and one
technological. Theoretically these gases are ideal for testing
fundamental superfluid properties (such as persistent currents,
for example), and they may realize exactly-solvable
one-dimensional models. Technologically, the possibility of
creating persistent currents whose stability is tunable
externally may lead to important, or even revolutionary
applications.

Motivated by these facts, we focus in the present study on the
stability of persistent currents in a quasi-one dimensional,
toroidal trap. Three are the classes of problems that we
consider. In the first one, we consider the situation where one
manipulates the trap appropriately, in order to achieve the
desired effect on the stability of currents. In the second, we
investigate unavoidable complications in realistic experiments,
which may affect the stability of persistent currents. The
third class includes the problems associated with the detection
of persistent currents.

Many aspects of the problem that we examine in our study have
been investigated thoroughly in previous studies. We do not
attempt to give a complete list of references, but rather we
mention just few of them
\cite{pl,mf,Huang,Giorgini1,Hakim,Pavloff,Giorgini2,Tsubota,Graham,LSP,Wiersma1,Wiersma2,Bouyer1,Bouyer2,Paul}.

In what follows, we first consider our model in Sec.\,II. Then,
in Sec.\,III we consider a step potential along the torus and
investigate the conditions that destabilize the current most
easily. In Sec.\,IV we consider a periodic potential along the
torus and examine the stability of persistent currents in such
a potential. Then, in Sec.\,V we consider a random potential
and via a statistical analysis of our results, we examine the
effect of the length scale of the irregularities on the
stability of the current. In Sec.\,VI, we investigate the
effect of gravity on a tilted torus, and in Sec.\,VII we
propose a method which allows the detection of a current.
Finally, in Sec. VIII we present the conclusions of our study.

\section{Model}

In our model we consider one-dimensional motion of the atoms
with periodic boundary conditions, which corresponds physically
to the motion of the atoms in a very tight toroidal trap.
Quasi-one dimensional motion is accomplished as long as the
interaction energy is much smaller than the excitation energy
transversely to the axis of the torus. We also assume for the
atom-atom collisions the usual contact potential, $V_{\rm
int}({\bf r}-{\bf r}') = U_0 \delta({\bf r}-{\bf r}')$ with
$U_0 = 4 \pi \hbar^2 a_{\rm sc} /M$. Here, $a_{\rm sc}$ is the
scattering length for elastic atom-atom collisions and $M$ is
the atomic mass.

The mean-field approximation that we use in this study implies
that the condensate order parameter $\Psi(\theta)$ satisfies
the nonlinear, Gross-Pitaevskii-like equation \cite{Ueda,GMK},
\begin{eqnarray}
  - \frac {\partial^2 \Psi} {\partial \theta^2}
  + V(\theta) \Psi(\theta)
  + 2 \pi \gamma|\Psi(\theta)|^2 \Psi
  = \mu \Psi,
\label{gpee}
\end{eqnarray}
where we have set $\hbar = 2M = R = 1$, with $R$ being the
radius of the torus. Here $\theta$ is the azimuthal angle,
$V(\theta)$ is the external potential, and $\mu$ is the
chemical potential. Both $V(\theta)$ and $\mu$ are measured in
units of the kinetic energy $T = \hbar^2/(2 M R^2)$. The ratio
between the interaction energy and the kinetic energy is equal
to $\gamma = n_0 U_0 / T = 4 N a_{\rm sc} R/S$. Here $n_0 = N
/(2 \pi R S)$ is the average atom density, $N \gg 1$ is the
atom number, and $S$ is the cross section of the torus (with $R
\gg \sqrt S$). As long as $\gamma \ll N^2$, the system is away
from the Tonks-Girardeau limit of impenetrable bosons
\cite{KYOR}, that we do not consider in our study.

To investigate the stability of persistent currents, we use the
time-dependent version of Eq.\,(\ref{gpee})
\begin{eqnarray}
  i \hbar \frac {\partial \Psi} {\partial t}  =
  - \frac {\partial^2 \Psi} {\partial \theta^2}
  + V(\theta) \Psi(\theta)
  + 2 \pi \gamma|\Psi(\theta)|^2 \Psi.
\label{gpeett}
\end{eqnarray}
We propagate some initial state in imaginary time, making the
substitution $\tau = i t$ \cite{imag}, and solve the equation
\begin{eqnarray}
  - \hbar \frac {\partial \Psi} {\partial \tau}  =
  - \frac {\partial^2 \Psi} {\partial \theta^2}
  + V(\theta) \Psi(\theta)
  + 2 \pi \gamma|\Psi(\theta)|^2 \Psi -\mu \Psi.
\label{gpeet}
\end{eqnarray}
More specifically, in the above equation we specify
$\Psi(\theta, \tau=0)$ and look for the convergent solution
that emerges for long enough times from the above
time-dependent equation.

The states that correspond to successive values of (the
quantized) circulation may be separated by an energy barrier,
provided that the interaction is strong enough
\cite{Bl,Leggett,KYOR}. These barriers give rise to stable
persistent currents. Therefore, if one starts with the initial
condition $\Psi(\theta, \tau = 0) = e^{i \theta} /\sqrt{2 \pi}$
(for example) that has one unit of circulation and a uniform
density distribution, if there exists an energy barrier between
this state and the current-free state, then the long-time
solution of Eq.\,(\ref{gpeet}), still has one unit of
circulation, with possibly some angle-dependent variation in
its density, depending on the form of the potential
$V(\theta)$. On the other hand, in the absence of an energy
barrier between the two states (which is the case for
sufficiently weak, or attractive interactions) the system is
energetically unstable, and it relaxes to the circulation-free
state.

The criterion for the stability of circulation that we use in
our study is equivalent to a condition that resembles the
Landau criterion for superfluid flow \cite{Bl}. More precisely,
in order to have stability of the superflow, the ``drift"
velocity of the atoms $u$ must not exceed the speed of sound
$c$. For example, for one unit of circulation that we consider
here, the drift velocity is equal to $\hbar/(MR)$. Also, in the
limit of weak interactions the speed of sound is given by
\cite{sound} (see, e.g., Eq.\,(\ref{varenergy}), or
Ref.\,\cite{OK}),
\begin{eqnarray}
   c = \frac {\hbar} {2 M R} (1 + \gamma),
\end{eqnarray}
and in order for $u$ to be smaller than $c$, $\gamma$ has to be
larger than unity, in agreement with the criterion of energetic
stability. Finally, we mention that while we have restricted
our study to the transition between states of circulation equal
to unity and zero, similar effects occur in transitions between
states of higher, successive values of the circulation.

\section{Destabilization of persistent currents with a step potential}

As a first application of our method, we consider a step-like
potential of a fixed height (equal to unity in our units) along
the torus,
\begin{equation}
V(\theta) = \cases{1, -\pi \le \theta \le \theta_0 \cr 0,
\theta_0 < \theta < \pi},
\label{steppot}
\end{equation}
and investigate the critical value $\gamma_c$ of the coupling
$\gamma$ that is necessary to maintain the stability of the
current, as function of $\theta_0$. The form of the potential
that we choose is not accidental. Among all the possible
functional forms of $V(\theta)$ that we investigated with
$V_{\rm max}-V_{\rm min}=1$, this is the one that requires the
highest value of $\gamma_c$, for a static potential. Therefore
under these conditions, at least within the functional forms of
the potentials that we have considered, the derived value of
$\gamma_c$ may serve as an upper bound for the critical
coupling that is necessary to sustain a persistent current of
circulation equal to unity.

Following the method that was described in the previous
section, we plot in Fig.\,1 $\gamma_c$ versus $\theta_0$. As
one sees in this plot, the highest value of $\gamma_c$ that is
necessary to stabilize the current occurs for $\theta_0/(2 \pi)
\simeq -0.18$. The position of $\theta_0$ decreases with
increasing $V_{\rm max} - V_{\rm min}$.

For $\theta_0 \to -\pi$, and $\theta_0 \to \pi$, $\gamma_c$
tends to $3/2$, which is the critical value of $\gamma_c$ that
corresponds to a constant $V(\theta)$ \cite{Ueda,GMK,OK}. For
intermediate values of $\theta_0$ the situation becomes more
interesting, as in this case the length scale of variation of
$V(\theta)$ is comparable to the coherence length $\xi$ and as
a result $\gamma_c$ is higher than 3/2. More precisely, when
$\gamma$ is of order unity, the coherence length $\xi$, is
$\sim R$, since $\xi/R = \gamma^{-1/2}$. The fact that $\xi$ is
comparable with the radius of the torus, and thus comparable
with the length scale of variation of $V(\theta)$, implies that
the current becomes fragile. A crude guess for $\theta_0/(2
\pi)$ that requires the maximum value of $\gamma_c$ is the
middle of the torus (i.e., when the step potential extends over
one half of the torus), as we also discuss in the toy model
that follows below. We mentioned earlier that for $V_{\rm max}
- V_{\rm min}=1$ the actual value is $\simeq -0.18$, with the
corresponding value of $\gamma_c$ being $\simeq 3.8$.

A simple toy model for this problem gives a qualitatively
correct answer for $\gamma_c$ as function of $\theta_0$. Let us
consider just the states $\Phi_0(\theta) = 1/\sqrt{2 \pi}$ and
$\Phi_1(\theta) =e^{i \theta}/\sqrt{2 \pi}$ in the order
parameter, in the form
\begin{equation}
   \Psi_{\rm trial}(\theta) = \sqrt{1 - l} \, \Phi_0(\theta)
   + e^{i \lambda} \sqrt l \, \Phi_1(\theta),
\end{equation}
where $\lambda$ is some phase factor. The value of $\lambda$ is
determined from the minimization of the potential energy, which
turns out to be $\lambda = - (\theta_0 + \pi)/2$. Clearly the
above state has an expectation value of the angular momentum
per particle equal to $l$. The corresponding density is
\begin{equation}
  n_{\rm trial}(\theta) = \frac 1 {2 \pi} [1 + 2 \sqrt{l (1-l)}
  \sin(\theta - \theta_0/2)],
\label{denprof}
\end{equation}
and also the (minimized) energy per particle is, for the step
potential of Eq.\,(\ref{steppot}),
\begin{equation}
   E - \frac \gamma 2 - \frac {\theta_0 + \pi} {2 \pi} =
   (1 + \gamma) l - \gamma l^2 - \sqrt {l (1-l)}
   \frac 2 \pi \cos( \frac {\theta_0} 2).
\label{varenergy}
\end{equation}
From the above dispersion relation we find the critical value
$\gamma_c$ that gives rise to a local minimum close to $l=1$.
The specific (highly truncated) order parameter gives $\gamma_c
= 1$ for a constant potential (i.e., for $\theta_0 = \pm \pi$),
while $\gamma_c$ has a maximum for $\theta_0 = 0$, equal to
$\approx 2.295$. In addition, $\gamma_c(\theta_0)$ is symmetric
around $\theta_0 = 0$. In the limit of weak disorder and weak
coupling, where the actual order parameter is closer to
$\Psi_{\rm trial}(\theta)$, the density profile of the cloud is
sinusoidal, and the step potential just determines the position
of the maximum/minimum of the density, as one can see in
Eq.\,(\ref{denprof}). As a result, $\gamma_c(-\theta_0) =
\gamma_c(\theta_0)$. For higher couplings and higher values of
$V_{\rm max} - V_{\rm min}$, the density of the cloud gets
distorted from the sinusoidal form, as it localizes more in the
region of lower potential energy. The function
$\gamma_c(\theta_0)$ becomes then asymmetric with respect to
$\theta_0 = 0$, as shown in Fig.\,1 (which shows the full
numerical calculation.)

\begin{figure}[t]
\includegraphics[width=8cm,height=5.0cm]{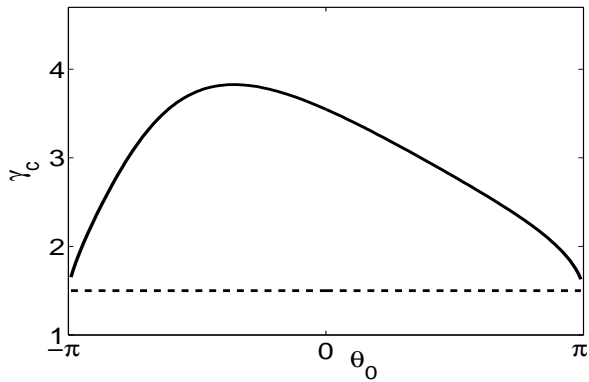}
\caption[]{The critical value of the coupling constant
$\gamma_c$ versus the location of the step $\theta_0$, in a
step potential of the form of $V(\theta)$ defined in
Eq.\,(\ref{steppot}). The value of $\gamma_c=3/2$ for the
uniform case (dashed line) is also shown as a reference.}
\label{FIG1}
\end{figure}

\section{Stability of persistent currents in the presence of a
periodic potential}

Another interesting question is the stability of persistent
currents in the presence of a periodic potential that acts
along the torus. Experiments with periodic potentials created
by optical lattices have already been performed \cite{optlat}.
Numerous theoretical studies have examined this problem, too.
This is a very interesting problem, since the presence of a
periodic external potential, combined with the effect of the
interactions give rise to novel states. For example, we refer
to Refs.\,\cite{Wu1,Bronski,Wu2,Diakonov,Machholm} for studies
of the Gross-Pitaevskii equation in the presence of a periodic
potential, for calculations of the band structure, and for the
study of the superfluid properties of these systems.

In our study we consider a sinusoidal external potential of the form
\begin{equation}
  V(\theta) = \cos (m \theta),
\label{perpot}
\end{equation}
where $m = 1, 2, 3, \dots$, as the condition $V(\theta + 2 \pi)
= V(\theta)$ requires. This potential has a period equal to $2
\pi/m$. Again, we examine the stability of a current with one
unit of circulation, starting with $\Psi(\theta, \tau = 0) =
e^{i \theta}/\sqrt{2 \pi}$ and finding the critical value of
$\gamma$ that gives rise to a stable current, for various
values of $m = 1, \dots, 6$. The results of this calculation
are shown in Fig.\,2. As $m$ increases, the value of $\gamma_c$
that is required to give rise to a stable current decreases.

For $m=1$, the potential is qualitatively similar to a step
potential examined in Sec.\,III, but it is smoother (which
decreases $\gamma_c$) and has a larger $V_{\rm max} - V_{\rm
min}=2$ (which increases $\gamma_c$). For a step potential with
$V_{\rm max} - V_{\rm min}=2$, that is located in the middle of
the torus ($\theta_0=0$), one finds $\gamma_c \simeq 4.8$, in
rough agreement with the maximum value of $\gamma_c$ shown in
Fig.\,2. Furthermore, as $m$ increases, eventually the
potential becomes homogeneous (on the length scale of variation
of the density), in which case $\gamma_c = 3/2$
\cite{Ueda,GMK,OK}. This is clearly seen in Fig.\,2. The
crossover region to the homogeneous case is accomplished when
the ``lattice constant" $\pi/m$, associated with the periodic
potential $V(\theta)$, is $\sim \xi/R = \gamma_c^{-1/2}$. The
above equation implies that $m \sim \pi \gamma_c^{1/2} = \pi
(3/2)^{1/2} \approx 3.9$, in (rough) agreement with our result
shown in Fig.\,2.

\begin{figure}[t]
\includegraphics[width=8cm,height=5.0cm]{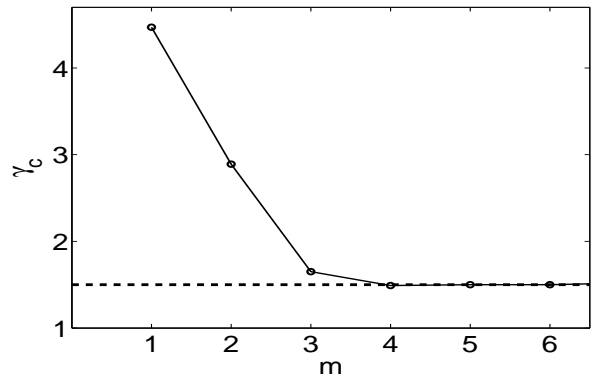}
\caption[]{The critical value of $\gamma_c$ versus the
parameter $m$ (dots), for the periodic potential of
Eq.\,(\ref{perpot}). The solid line is a guide for the eye. The
value of $\gamma_c=3/2$ for the uniform case (dashed line) is
also shown as a reference.}
\label{FIG2}
\end{figure}

\section{Stabilization of a current in the presence of a random potential}

In a previous study \cite{OK}, we investigated the value of
$\gamma_c$ that is necessary to stabilize a current in the
presence of a piecewise constant potential, with a randomly
chosen amplitude. Here we present data that we have collected
from a statistical analysis of our simulations, where we
examine $\gamma_c$ as function of the number of steps that we
choose in $V(\theta)$.

More specifically, we compute the average value of $\gamma_c$,
as well as the standard deviation $\sigma(\gamma_c)$, for 1000
different random step potentials, for which there are $s$
steps, with a width $(2 \pi/s)$. The value of the potential
within each step is drawn uniformly from the interval $[-1,
1]$. Figure 3 shows the result of these calculations. We
observe that $\langle \gamma_c \rangle$ is a decreasing
function of $s$.

As in the case of a step potential, one may argue that as $s$
increases, the length scale of variation of the random
potential becomes increasingly smaller than the coherence
length, and eventually one recovers the result $\gamma_c=3/2$
of the homogeneous torus, when $V(\theta)$ is constant. On the
other hand, as $s$ decreases, eventually the two length scales
become comparable and the random potential destabilizes the
current more easily, requiring a higher value of $\gamma_c$ for
the current to become stable.

\begin{figure}[t]
\includegraphics[width=8cm,height=5.0cm]{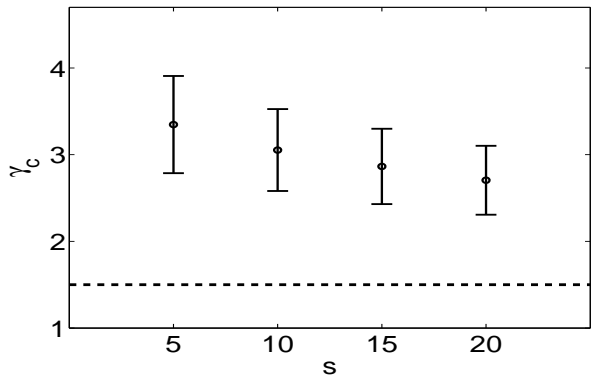}
\caption[]{The average value of the critical coupling $\langle
\gamma_c \rangle$ (dots), as well as the standard deviation
(bars), versus the number of steps $s$ in a randomly-chosen
potential. These results are derived from 1000 different random
potentials for each number of steps $s=5, 10 ,15, 20$. The
value of $\gamma_c=3/2$ for the uniform case (dashed line) is
also shown as a reference.}
\label{FIG3}
\end{figure}

\section{Effect of gravity on the stability of persistent currents}

One question that needs to be investigated within the problems
that we examine here is the effect of gravity on the stability
of persistent currents. Already in other experiments with cold
atoms, gravity has played an important role, see e.g.
\cite{Mewes}. In real life, the torus cannot be perfectly
horizontal, and as a result, there is an angle-dependent
potential that acts on the atoms due to the gravitational
force. This potential is
\begin{equation}
   V(\theta) = (M g R \sin \delta) \cos \theta,
\label{gravity}
\end{equation}
where $g$ is the acceleration of gravity and $\delta$ is the
angle between the plane of the torus and the horizontal plane.

Figure 4 shows $\gamma_c$ as function of $\log_{10} (\delta)$,
for three values of $R = 1$ mm (left), 0.1 mm (middle), and
0.01 mm (right). As the angle $\delta$ increases for a given
$R$, $\gamma_c$ increases too, since the effect of the
gravitational field gets more important. For the same reason,
for a fixed angle $\delta$, as the radius of the torus $R$
increases, the critical value of $\gamma_c$ increases, too. For
the small values of the angle $\delta$ that we have considered,
$\sin \delta \approx \delta$, and therefore the shape of the
three different curves that are shown (which correspond to the
three different values of $R$), are almost identical, as they
can be obtained approximately by a shift in $\delta$, which is
$\propto R^{-3}$.

It is instructive to get an estimate for the energy scale $M g
R$ in Eq.\,(\ref{gravity}). Considering Rubidium atoms, for
example, and for a radius of the torus $R = 0.1$ mm, $M g R$ is
on the order of $\mu$K, which is a rather large energy scale.
Therefore, in order for the effect of gravity to be
unimportant, the torus has to be tilted slightly, in which case
$\sin \delta \ll 1$.

\begin{figure}[t]
\includegraphics[width=8cm,height=5.0cm]{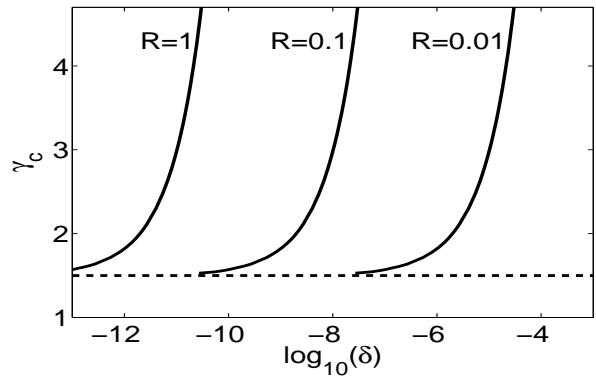}
\caption[]{The critical value of $\gamma_c$ versus $\log_{10}
(\delta)$, for $R = 1$ mm (left), 0.1 mm (middle), and 0.01 mm
(right). The value of $\gamma_c=3/2$ for the uniform case
(dashed line) is also shown as a reference.}
\label{FIG4}
\end{figure}

\section{Detection of persistent currents}

A serious issue in these experiments is to be able to know
whether there is circulation in the gas, or not. This question
can be resolved with use of interference techniques
\cite{Chevy}, however it would be convenient to have also a way
to measure the circulation using other methods.

It is natural to think of the single-particle density
distribution $n(\theta) = |\Psi(\theta)|^2$ as a possible way
to determine the value of the circulation. Any constant
potential $V(\theta)$ results in a homogeneous density
distribution, as the order parameter is $e^{i \kappa \theta}
/\sqrt{2 \pi}$, for any value of the circulation $\kappa (2 \pi
\hbar/M)$. On the other hand, if one uses a probe potential
$V(\theta)$ that is spatially-dependent, the density
distribution $n(\theta)$ does depend on the value of $\kappa$.
Therefore, measuring the density in the presence of some
spatially-dependent potential $V(\theta)$ may allow us to
determine the value of the circulation.

We thus consider such a probe potential in the form of a
Gaussian unity dimple (which may be realized via some laser
beam) with (rms) width $w_d$
\begin{equation}
  V(\theta)  =  -\exp(-\theta^2/2w_d^2).
\label{probe}
\end{equation}
Figure 5 shows the density distribution $n(\theta)$ in the
presence of $V(\theta)$ of Eq.\,(\ref{probe}), with $w_d=0.2$
for the states with zero circulation ($\kappa = 0$), and unit
circulation ($\kappa = 1$). The chosen value of the coupling is
larger than the critical one, $\gamma=2.2 > \gamma_c \approx
2.1$. The state with $\kappa=1$ has a higher kinetic energy,
and as a result the classically-forbidden region is more
narrow, resulting in a wider density distribution around the
minimum of the probe potential $V(\theta)$. Ideally one would
like the density profiles corresponding to different values of
the circulation to be as different as possible, even in the
limit of a very weak probe potential $|V(\theta)| \ll 1$.

To quantify the difference between the two density
distributions illustrated in Fig. 5, one may introduce the
difference between the highest and the lowest value of the
density in each case, $\Delta n = n_{\rm max} - n_{\rm min} =
n(\theta = 0) - n(\theta = \pm \pi)$. Remarkably, this quantity
differs by more than a factor of two ($\approx 2.7$) between
the two density profiles that we consider in the specific
example.

The above results indicate that a possible way to measure the
circulation (non-destructively), would be to turn on the probe
potential adiabatically, measure the density and extract the
value of $\kappa$ (non-destructively), and finally turn off the
probe adiabatically again.

\begin{figure}[t]
\includegraphics[width=8cm,height=5.0cm]{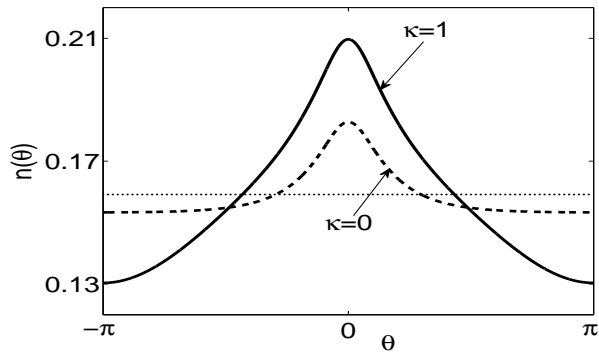}
\caption[]{The (normalized) density distribution $n(\theta)$ of
the state of lowest energy, in the presence of $V(\theta)$
given by Eq.\,(\ref{probe}), for zero circulation, $\kappa = 0$
(dashed curve), and unit circulation $\kappa = 1$ (solid
curve), for a coupling strength $\gamma=2.2$. The horizontal
dotted line shows the density of the uniform state,
$n(\theta)=1/2\pi$. All densities are measured in units of
$N/(RS)$.}
\label{FIG5}
\end{figure}

\section{Summary and conclusions}

In this study we examined the behavior of atoms that circulate
in a tight toroidal trap. Manipulation of the trapping
potential along the torus may provide various ways to control
the stability of the current. The form of the potentials that
we considered include a step potential, a sinusoidal potential,
a random potential, and a gravitational potential. We
calculated the value of the critical coupling that gives rise
to stable persistent currents. In addition, we suggested a
method that allows the detection of circulation. This method
involves a probe, spatially-dependent potential, which, for
different values of the circulation gives a different density
variation. This method requires only a rough measurement of the
density profile, which may be realizable non-destructively
\cite{Brennecke}.

Our results indicate that the remarkable progress on the
manipulation of the trapping potential \cite{Schnelle}, and
more generally the developments in the physics of cold atoms
may allow us to engineer such mesoscopic systems that support
persistent currents according to our will. Such systems may
therefore serve as ``superconducting switches" with possible
important technological applications.

\acknowledgments We thank Jozsef Fortagh and Andrew Sykes for
useful discussions.


\begin{thebibliography}{99}
\bibitem{sound} This result implies that $M c^2 = \hbar^2/(4 M R^2)
+ 2 \hbar^2 N a_{\rm sc}/(MRS) + {\cal O}(\gamma^2)$. In the
limit $N \to \infty$ and $R \to \infty$, with $N/R$ finite, $M
c^2 = 2 \hbar^2 N a_{\rm sc}/(MRS)$, in agreement with the
formula $M c^2 = n_0 U_0$, which is known for a homogeneous gas
of density $n_0 = N/(2 \pi R S)$.
\end{thebibliography}
\end{document}